\documentclass[12pt,twoside]{article}   
\usepackage[super,sort,comma]{natbib}

\usepackage{fancyhdr}		
\usepackage{amssymb}
\usepackage{amsmath,amscd}
\usepackage{bm}
\usepackage{verbatim} 

\usepackage[section]{placeins}   %

\usepackage{graphicx}

\makeatletter \renewcommand\@biblabel[1]{$^{#1}$} \makeatother
\newcommand{\cen}[1]{\begin{center} #1 \end{center}}

\usepackage[mathlines]{lineno}

\usepackage{hyperref}
\hypersetup{ colorlinks,
    citecolor=blue,
    filecolor=blue,
    linkcolor=blue,
    urlcolor=blue
}

\usepackage{xcolor}

\definecolor{gray}{rgb}{0.6,0.6,0.6}
\definecolor{red}{rgb}{0.85,0,0}
\definecolor{green}{rgb}{0,0.85,0}
\definecolor{blue}{rgb}{0,0,0.85}
\definecolor{beige}{rgb}{0.92,0.87,0.78}
\usepackage[all]{hypcap}    
\usepackage{booktabs} 
\usepackage{float} 
\usepackage{booktabs}  
\usepackage{multirow} 
\usepackage{rotating}  
\usepackage{siunitx} 
\usepackage{caption} 
\usepackage{xcolor} 
\definecolor{resultred}{RGB}{200,50,50} 
\begin{document}

\cen{\sf {\Large {\bfseries A Comprehensive Scatter Correction Model for Micro-Focus Dual-Source Imaging Systems: Combining Ambient, Cross, and Forward Scatter} \\  

\vspace*{10mm}
\normalsize Jianing Sun\textsuperscript{1,2}\hspace{1mm}, Jigang Duan\textsuperscript{1,2}\hspace{1mm}, Guangyin Li\textsuperscript{1,2}\hspace{1mm}, Xu Jiang\textsuperscript{1,2}\hspace{1mm}, Xing Zhao\textsuperscript{1,2}$^{*}$} \\[5mm]
{\footnotesize 
\textsuperscript{1} School of Mathematical Sciences, Capital Normal University, Beijing 100048, China,\\
\textsuperscript{2} Beijing Higher Institution Engineering Research Center of Testing and Imaging, Beijing, China \\
}
\vspace{5mm}
{\footnotesize 
$^{*}$\text{Corresponding Author:} Xing Zhao \\
\text{Email:} zhaoxing\_1999@126.com \\
}
}
\pagenumbering{roman}
\setcounter{page}{1}
\pagestyle{plain}

\begin{abstract}
Compared to single-source imaging systems, dual-source imaging systems equipped with two cross-distributed scanning beams significantly enhance temporal resolution and capture more comprehensive object scanning information. 
Nevertheless, the interaction between the two scanning beams introduces more complex scatter signals into the acquired projection data.
Existing methods typically model these scatter signals as the sum of cross-scatter and forward scatter, with cross-scatter estimation limited to single-scatter along primary paths.
Through experimental measurements on our self-developed micro-focus dual-source imaging system, we observed that the peak ratio of hardware-induced ambient scatter to single-source projection intensity can even exceed $60\%$, a factor often overlooked in conventional models. 
To address this limitation, we propose a more comprehensive model that decomposes the total scatter signals into three distinct components: ambient scatter, cross-scatter, and forward scatter. 
Furthermore, we introduce a cross-scatter kernel superposition (\textbf{xSKS}) module to enhance the accuracy of cross-scatter estimation by modeling both single and multiple cross-scatter events along non-primary paths.
Additionally, we employ a fast object-adaptive scatter kernel superposition (\textbf{FOSKS}) module for efficient forward scatter estimation. 
In Monte Carlo (MC) simulation experiments performed on a custom-designed water-bone phantom, our model demonstrated remarkable superiority, achieving a scatter-to-primary-weighted mean absolute percentage error (SPMAPE) of $1.32\%$, significantly lower than the $12.99\%$ attained by the state-of-the-art method. Physical experiments further validate the superior performance of our model in correcting scatter artifacts. 
\end{abstract}



\newpage

\setlength{\baselineskip}{0.7cm}      

\pagenumbering{arabic} 
\setcounter{page}{1}
\pagestyle{fancy} 
\section{Introduction}
The dual-source imaging system, employing two sets of cross-distributed scanning beams, offers significant advantages over the single-source configuration. Particularly, it enhances temporal resolution and mitigates motion artifacts in cardiac imaging \cite{flohr2008image}. Furthermore, by operating the two scanning beams at different voltages and utilizing distinct filters, the system enables simultaneous acquisition of multi-energy projections at both high and low energies. This capability facilitates the direct reconstruction of base material images and pseudo-monoenergetic images, effectively eliminating beam-hardening artifacts and significantly improving quantitative identification performance \cite{pan2023fast}.

However, the dual-source configuration introduces additional challenges beyond the forward scatter typically observed in single-source systems. Specifically, the dual-source design gives rise to cross-scatter, a distinct phenomenon resulting from the interference between projections acquired by the two scanning beams. Both forward scatter and cross-scatter distort primary transmission, with the intensity of cross-scatter in small objects being comparable to that of forward scatter and increasing proportionally with object size \cite{kyriakou2007intensity}. 
Moreover, the superposition of forward scatter and cross-scatter can even reach twice the intensity of primary transmission \cite{engel2008x}.
The combined effects of these scattering mechanisms lead to significant scatter artifacts in reconstructed images, which degrade image contrast and compromise the accuracy of quantitative detection.

While scatter correction methods for single-source imaging systems are well-established, encompassing a wide range of hardware-based solutions (e.g., anti-scatter grids \cite{siewerdsen2004influence} and air gaps \cite{ruhrnschopf2011general}) and software-based approaches (e.g., Monte Carlo simulations\cite{badal2009accelerating,lin2021quasi} (MC), deep learning techniques\cite{maier2018deep,jiang2022generalized,zhang2023image,zhuo2023scatter} and mathematical model\cite{sun2010improved,mason2018quantitative}), scatter correction for dual-source systems remains significantly more challenging. Hardware-based solutions, such as anti-scatter grids and air gaps, have shown limited effectiveness in suppressing cross-scatter\cite{engel2008x, kyriakou2007intensity, gong2017physics}. Although grid blocks have been proposed for scatter estimation \cite{ren2016scatter}, this approach requires manual adjustments and multiple scans, significantly increasing operational complexity. A Sensor-based method\cite{petersilka2010strategies} and a beam-stopper-array (BSA) method\cite{gong2017x}, while promising, introduce additional hardware complexity and implementation challenges. 

Software-based correction methods, which do not rely on specialized hardware, primarily leverage deep learning techniques and mathematical models. Among the deep learning approaches, the deep scatter estimation (DSE) network, based on the U-Net architecture, has emerged as a pioneering solution. It has demonstrated significant effectiveness not only in estimating scatter in single-source setups \cite{maier2018deep}, but also in predicting forward scatter in dual-source configurations \cite{erath2021deep}. Extensions of this approach, including the xDSE and xDSE3D networks, have shown promise in generating cross-scatter estimates that align closely with ground truth data in dual-source systems \cite{erath2021deep}. However, these methods require extensive training data and computational resources, posing significant challenges for specific detection tasks. Among mathematical models, the lookup table approach characterizes surface properties with three variables to estimate  cross-scatter \cite{petersilka2010strategies}, but its reliance on diverse objects for table generation makes it cumbersome. The single scatter estimation for cross-scatter (xSSE) algorithm, used to improve the training accuracy of the xDSE network, calculates the single cross-scatter along primary paths \cite{erath2021deep}. While it provides an initial estimate of cross-scatter, it neglects scatter contributions along non-primary paths, which significantly degrade the reconstructed image quality. The Klein-Nishina (KN) formula-based iterative method \cite{gong2017physics}, which is applicable to multi-source scatter correction, effectively reduces scatter artifacts but incurs high computational costs even with a few iterations. Another affine projection correction protocol (APCP) algorithm  for micro dual-source systems models total scatter as a combination of object-induced scatter (including both forward and cross-scatter) and detector-induced cross-scatter \cite{letang2024cross}. This method assumes that object-induced scatter remains constant but varies with the scan angle, while the shape of detector-induced cross-scatter remains invariant from the air-scan condition, with its magnitude varying as a constant at different rotation angles. This method is particularly effective at high magnification ratios as scatter smoothness increases with air gap size\cite{boellaard1997convolution}.

Despite these advancements, existing methods have yet to fully address the impact of hardware-induced scatter during the scanning process, referred to as ambient scatter. Through experimental measurements of phantoms conducted on our self-developed micro-focus dual-source imaging system, we observed that the peak ratio of ambient scatter to single-source projection intensity can even exceed $60\%$, a factor frequently overlooked in conventional models. To address this limitation, we propose a comprehensive model that decomposes total scatter signals into three distinct components: ambient scatter, cross-scatter, and forward scatter. 
Firstly, we establish that angular variations in ambient scatter have negligible impact on the accuracy of reconstructed images, enabling it to be modeled as a discrete point source and measured during an air-scan. 
Furthermore, we propose the xSKS module to enhance the accuracy of cross-scatter estimation by modeling both single and multiple cross-scatter events accuring along non-primary paths, which are neglected in the xSSE algorithm.
Specifically, this module conceptualizes cross-scatter as a convolution of single cross-scatter from primary paths and a double Gaussian kernel. Additionally, by incorporating measured prior knowledge of cross-scatter, the parameters of the xSKS module are pre-calibrated, facilitating rapid and efficient cross-scatter estimation at any scanning angle. In contrast to parameter calibration relying on MC simulations, the use of experimentally acquired priors improves the accuracy of cross-scatter correction under physical scanning conditions. 
Finally, extending the fast adaptive scatter kernel superposition (FASKS) algorithm \cite{sun2010improved}, which estimates scatter in single-source imaging systems, we introduce the FOSKS module to correct forward scatter in micro-focus dual-source configurations. This module leverages forward scatter priors obtained from MC simulations at sparse angles to pre-calibrate its parameters, eliminating the limitations associated with using water slabs of varying thicknesses.

In summary, the contributions of this work are outlined as follows. 
First, we propose a comprehensive model that decomposes total scatter signals into three distinct components: ambient scatter, cross-scatter, and forward scatter. 
To the best of our knowledge, this represents the first attempt to model scatter in dual-source imaging systems as the sum of these three components.
Second, we introduce a novel analytical xSKS module, which extends the capabilities of the existing xSSE algorithm by further estimating the cross-scatter along non-primary paths.
Third, we utilize the FOSKS module to address forward scatter correction, which overcomes the limitations of the existing FASKS algorithm in parameter calibration. 
Collectively, these contributions advance the state-of-the-art in scatter correction for dual-source imaging systems, offering a more efficient and practical solution for improving image quality and quantitative accuracy.

\section{Methods}
\subsection{Proposed Scatter Model}
In the conventional single-source imaging system, the  measured projection intensity $I$ consists of two components: the primary transmission $I^{p}$ and the forward scatter $I^{s}$, which can be expressed as:
\begin{align}
\label{single-total}
   I\{k\}(u,v) = I^{p}\{k\}(u,v) + I^{s}\{k\}(u,v).
\end{align}
In this equation, the notation $\{k\}(u,v)$ denotes the detector coordinates $(u,v)$ corresponding to the $k$-th projection angle. According to the Lambert-Beer law, the mathematical formulation of the primary transmission $I^{p}$ is given by:
\begin{align}
I^{p}\{k\}(u,v) = I^{*}(u,v)\mathrm{exp}\left(-\int_{l\in L_k} \mu(\vec{x}) \mathrm{d}l\right),
\end{align} 
where $I^{*}$ represents the incident photon intensity obtained during the air-scan process, $L_{k}$ denotes the set of all ray paths from the X-ray source to the detector pixel $(u,v)$ at the $k$-th projection angle, and $\mu(\vec{x})$ is the linear attenuation coefficient at the position $\vec{x}$ within the object under examination.

The dual-source setup incorporates two sets of cross-distributed scanning beams. Consequently, the measured total projection signal is influenced not only by forward scatter but also by cross-scatter interference originating from the X-ray source in the orthogonal direction.
Furthermore, through comprehensive experimental measurements conducted on a micro-focus dual-source imaging system, this study demonstrates that scatter is also significantly affected by the hardware configuration during the scanning process, which has a non-negligible impact on the reconstructed image. Therefore, we propose a comprehensive model to estimate the scatter in micro-focus dual-source imaging systems.

\begin{figure}[ht]
   \begin{center}
   \includegraphics[width=\textwidth]{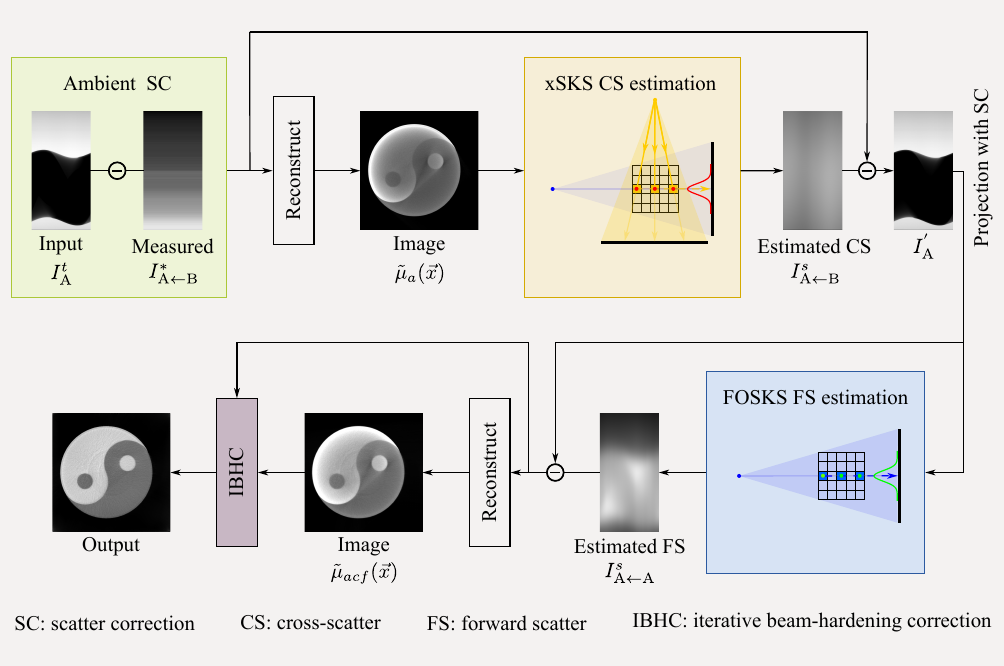}
   \caption{The flowchart illustrates our proposed scatter model, a novel framework designed to correct scatter artifacts in micro-focus dual-source imaging systems.}
   \label{tensorFlow11} 
    \end{center}
\end{figure}

Let the indices $i,j\in$ [A,B], where $i\neq j$. The total projection intensity $I_{i}^{t}$ of scanning beam $i$ is composed of four components:
\begin{align}
I_{i}^{t}\{k\}(u,v) = I_{i}^{p}\{k\}(u,v) + I_{i\leftarrow i}^{s}\{k\}(u,v) + I_{i\leftarrow j}^{s}\{k\}(u,v) + I_{i \leftarrow j}^{*}(u,v).
\end{align}
Among these components, $I_{i}^{p}$ represents the primary transmission of scanning beam $i$. $I_{i\leftarrow i}^{s}$ is the forward scatter from the X-ray source $i$ to scanning beam $i$. The $I_{i \leftarrow j}^{*}$ represents the ambient scatter from the X-ray source $j$ to scanning beam $i$. Taking scanning beam A as an example, the detected total projection intensity $I_{\text{A}}^{t}$ is:
\begin{align}\label{A-total}
   I_{\text{A}}^{t}\{k\}(u,v) = I_{\text{A}}^{p}\{k\}(u,v) + I_{\text{A}\leftarrow \text{A}}^{s}\{k\}(u,v) + I_{\text{A}\leftarrow \text{B}}^{s}\{k\}(u,v) + I_{\text{A}\leftarrow {\text{B}}}^{*}(u,v).
\end{align}
From Eq. \ref{A-total}, it is evident that when X-ray source A is deactivated while source B remains operational during an air-scan, both the primary transmission $I_{\text{A}}^{p}$ and forward scatter $I_{\text{A}\leftarrow \text{A}}^{s}$ vanish. Concurrently, $I_{\text{A}\leftarrow \text{B}}^{s}$, which depends on the linear attenuation coefficient of the object being measured, also approaches zero under air-scan. Under these circumstances, detector A captures the ambient scatter $I_{\text{A}\leftarrow {\text{B}}}^{*}$ from X-ray source B to beam A. Based on measurements of experimental phantoms conducted on the micro-focus dual-source imaging system, we determined that angular variations in ambient scatter have a negligible impact on the accuracy of reconstructed images. Consequently, we model the ambient scatter as a discrete point source, which can be measured at a single projection angle as described above (source B on, source A off).
Based on ambient scatter correction, we propose the xSKS module for cross-scatter correction and the FOSKS module for further forward scatter correction.
To mitigate potential beam-hardening artifacts, an iterative beam-hardening correction algorithm can be integrated into the reconstruction process.
The complete scatter correction process for micro-focus dual-source setups is illustrated in Fig. \ref{tensorFlow11}.

\subsection{xSKS module for Cross-Scatter}
\begin{figure}[H]
   \begin{center}
    \includegraphics[width=\textwidth]{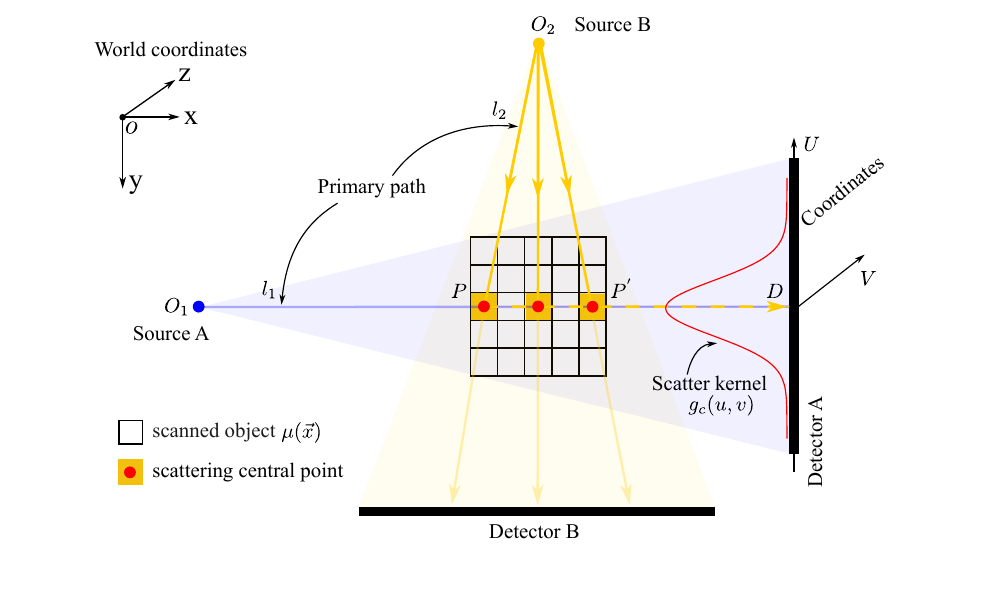}
   \caption{The schematic diagram of proposed xSKS module estimating cross-scatter.
   \label{xSKS_8} 
    }  
    \end{center}
\end{figure}

As illustrated in Fig. \ref{xSKS_8}, the X-ray photons emitted by X-ray source B traverse the measured object along the primary path $l_1$ of scanning beam A and undergo scattering at point $P$. According to Lambert-Beer's law, the primary transmission at point $P$ can be expressed as
\begin{align}
    I_{\text{B}}^{p}(P)\{k\}(u,v) = I^{*}(u,v)\,\mathrm{exp}\left(-\int_{l\in L_k} \mu(\vec{x}) \, \mathrm{d}x\right).
\end{align}
The cross-scatter $I^{s}(P)$, occurring along the path $\overrightarrow{PD}$ from point $P$, is strongly related to the linear attenuation coefficient $\mu(P)$:
\begin{align}
    I^{s}(P) =  k\,\mu(P)\,I_{\text{B}}^{p}(P),
\end{align}
where $k$ is the cross-scatter scaling factor to be calibrated. The scattered photons $I^{s}(P)$ undergo further attenuation along the path $\overrightarrow{PD}$ and are ultimately measured by detector unit $D$. When the scatter point $P$ traverses any point $P^{'}$ along the primary path $l_2$ of scanning beam A, the single cross-scatter from X-ray source B to the primary path $l_2$ is denoted as $S_{\text{A}\leftarrow\text{B}}$ :
\begin{align}\label{xSSE Method}
S_{\text{A}\leftarrow\text{B}}(D) = k\, I^{*}\int_{\overrightarrow{O_{1}D}} \exp\left(-\int_{\overrightarrow{O_{2}P^{'}}} \mu(\vec{x})\,\mathrm{d}l \right) \,\mu(P^{'}) \, \exp\left(-\int_{\overrightarrow{P^{'}D}} \mu(\vec{x})\, \mathrm{d}l\right) \mathrm{d}U^{'}.
\end{align}
Simultaneously, a non-negligible portion of the cross-scatter deviates from the primary path $l_2$ due to single or multiple scattering interactions, subsequently reaching various detector elements within detector A. To accurately model this phenomenon, the proposed xSKS module formulates the cross-scatter  $I_{\text{A}\leftarrow \text{B}}^{s}$ as the convolution of the single cross-scatter $S_{\text{A}\leftarrow\text{B}}$ with a cross-scatter kernel $g_c$, mathematically expressed as:
\begin{align}\label{xSKS}
I_{\text{A}\leftarrow \text{B}}^{s}\{k\}(u,v) = \iint\limits_{D_{\text{A}}} S_{\text{A}\leftarrow\text{B}}\{k\}(u^{\prime},v^{\prime})\,g_{c}(u-u^{\prime},v-v^{\prime})\,\mathrm{d}u^{\prime}\mathrm{d}v^{\prime},
\end{align}
where $D_{\text{A}}$ represents the coordinate plane of detector A. The cross-scatter convolution kernel $g_c$ is modeled as a superposition of two Gaussian functions:
\begin{align}\label{gc Kernel}
    g_{c}(u - u^{\prime},v - v^{\prime}) = \textup{exp}\left(-\frac{(u - u^{\prime})^2 + (v - v^{\prime})^2}{2{\tilde{\sigma}_1}^2}\right) + {\tilde{B}}\,\textup{exp}\left(-\frac{(u - u^{\prime})^2 + (v - v^{\prime})^2}{2{\tilde{\sigma}_2}^2}\right),
\end{align}
where ${\tilde{B}}$, ${\tilde{\sigma}_1}$, ${\tilde{\sigma}_2}$ denote the kernel parameters. Leveraging the equivalence between time-domain convolution and frequency-domain pointwise multiplication, cross-scatter $I_{\text{A}\leftarrow \text{B}}^{s}$ can be efficiently computed using the 2D fast Fourier transform (2D FFT):
\begin{align}\label{xSKS_FFT}
    I_{\text{A}\leftarrow \text{B}}^{s}\{k\}(u,v) = \mathcal{F}^{-1}\left(\mathcal{F}( S_{\text{A}\leftarrow\text{B}}\{k\}(u,v))\,\mathcal{F}\left(g_c(u,v)\right)\right).
\end{align}

Building on this foundation, the cross-scatter parameters of the xSKS module $\vec{\theta}_c =\left(k,\tilde{B},{\tilde{\sigma}_1},{\tilde{\sigma}_2}\right)$ are pre-calibrated through a systematic optimization process, leveraging prior knowledge derived from experimental measurements. 

Following the single-angle parameter calibration, the xSKS module is extended to estimate cross-scatter contributions across all projection angles. This allows deriving the ambient scatter and cross-scatter corrected projection intensity, denoted as $I_{\text{A}}^{\prime}$, using the following expression:
\begin{align}
    I_{\text{A}}^{\prime}\{k\}(u,v) &= I_{\text{A}}^{t}\{k\}(u,v) 
    - I_{\text{A} \leftarrow \text{B}}^{*} - \mathcal{F}^{-1} \left(\mathcal{F}(S_{\text{A} \leftarrow \text{B}}\{k\}(u,v)) \,
    \mathcal{F}(g_{c}(u,v)) \right) .
\end{align}
Despite the cross-scatter correction, the resulting projection intensity $I_{\text{A}}^{\prime}$  remain contaminated by forward scatter. Consequently, scattering artifacts persist in the reconstructed image $\tilde{\mu}_{ac}(\vec{x})$, necessitating an additional correction step to address forward scatter interference. 
\subsection{Forward Scatter Correction}
\subsubsection{Review of Scatter Kernel-Based Forward Scatter Correction}
In the scatter kernel superposition (SKS) method, the forward scatter $I_{\text{A} \leftarrow \text{A}}^{s}$ for scanning beam A is modeled as a convolution filtering process between the primary transmission $I_{\text{A}}^{p}$ and the scatter kernel $h_s$\cite{love1987scatter}. Specifically, it can be expressed as:
\begin{align}\label{sks}
    I_{\text{A}\leftarrow \text{A}}^{s}\{k\}(u,v) = \iint\limits_{D_{\text{A}}} I_{\text{A}}^{p}\{k\}(u^{\prime},v^{\prime})\, h_s\{k\}(u - u^{\prime},v - v^{\prime})\,\mathrm{d}u^{\prime} \mathrm{d}v^{\prime},
\end{align}
where $D_{\text{A}}$ is defined as in Eq. \ref{xSKS}, scatter kernel $h_s$ is composed of an amplitude factor $c$ and a double Gaussian function $g$, which can be formulated as:
\begin{align}\label{hs}
    h_s\{k\}(u - u^{\prime},v - v^{\prime}) = c\{k\}(u^{\prime},v^{\prime})\,g(u - u^{\prime},v - v^{\prime}).
\end{align}

\noindent The amplitude factor $c$ is determined by the primary transmission $I_{\text{A}}^{p}$ and the accident photon intensity $I^{*}$, and it can be expressed as:
\begin{align}
    c\{k\}(u^{\prime},v^{\prime}) = A\,\left(\frac{I_{\text{A}}^{p}\{k\}(u^{\prime},v^{\prime})}{I^{*}(u^{\prime},v^{\prime})}\right)^{\alpha}\,\left(\ln\left( \frac{I^{*}(u^{\prime},v^{\prime})}{I_{\text{A}}^{p}\{k\}(u^{\prime},v^{\prime})}\right)\right)^{\beta},
\end{align}
where $A, \alpha, \beta$ are parameters that influence the amplitude of the scatter kernel. The double Gaussian kernel function $g$ is defined as the sum of two symmetric Gaussian functions\cite{suri2006comparison}:
\begin{align}\label{guassKernel}
    g(u - u^{\prime},v - v^{\prime}) = \textup{exp}\left(-\frac{(u - u^{\prime})^2 + (v - v^{\prime})^2}{2\sigma_1^2}\right) + B\,\textup{exp}\left(-\frac{(u - u^{\prime})^2 + (v - v^{\prime})^2}{2\sigma_2^2}\right),
\end{align}
where $\sigma_1,\sigma_2$ represent the standard deviations of the two Gaussian functions, and $B$ is a weighting factor. The forward scatter parameters ($A,B,\alpha,\beta,\sigma_1,\sigma_2$) depend on the material properties and geometric shape of the measured object. Based on the thickness of the object through which the X-ray passes, the adaptive scatter kernel superposition (ASKS) algorithm calibrates these parameters \cite{sun2010improved}. The thickness group weight function $R^{i}(u,v)$ is defined as:
\begin{align}
    R^{i}(u,v) &= 
        \begin{cases}
                1,&   t_i \leq \tau(u,v)<t_{i+1}, i = 1\ldots n,\\ 
                0,&   \textup{otherwise} \\                         
       \end{cases}
\end{align}
where $n$ denotes the number of thickness interval groups, $t_i$ and $t_{i+1}$ represent the lower and upper bounds of the $i$-th thickness interval, respectively. The equivalent water thickness $\tau$ is expressed as: 
\begin{align}
   \tau\{k\}(u,v)\approx \frac{1}{\mu}\,\ln\left(\frac{I^{*}(u,v)}{I_{\text{A}}^{p}\{k\}(u,v)}\right),
\end{align}
where $\mu$ is the equivalent linear attenuation coefficient of single-material water at the scanning spectrum. Therefore, the forward scatter in the ASKS algorithm can be expressed as:
\begin{align}\label{ASKS}
I_{\text{A}\leftarrow \text{A}}^{s}\{k\}(u,v) &= \sum_{i=1}^n \iint\limits_D \mathcal{I}^{i}\{k\}(u^{\prime},v^{\prime}) \, g^{i}\{k\}(u - u^{\prime},v - v^{\prime})\,\mathrm{d}u^{\prime} \mathrm{d}v^{\prime}, \text{where} \nonumber \\
\mathcal{I}^{i}\{k\}(u^{\prime},v^{\prime}) &= I_{\text{A}}^{p}\{k\}(u^{\prime},v^{\prime}) \, R^{i}\{k\}(u^{\prime},v^{\prime}) \,c\{k\}(u^{\prime},v^{\prime}).
\end{align}
Based on the 2D FFT, FASKS algorithm represents the forward scatter as:
\begin{align}\label{FASKS}
   I_{\text{A}\leftarrow \text{A}}^{s}\{k\}(u,v) = \mathcal{F}^{-1} \sum_i \mathcal{F}\left(\mathcal{I}^{i}\{k\}(u,v)\right)\,\mathcal{F}(g^{i}\{k\}(u,v).
\end{align}

The FASKS algorithm offers an efficient approach for calculating forward scatter under single-source setups. While it offers fast computational performance and demonstrates the capability to correct forward scatter artifacts to a certain extent, several limitations remain to be addressed. Firstly, the forward scatter is inherently material-dependent. The current approach, which relies on forward scatter derived from water plates of varying thickness intervals for scatter parameter calibration, not only increases computational complexity but also tends to overestimate forward scatter in regions containing high-attenuation materials \cite{zhuo2023scatter}. This limitation highlights the need for a more robust material-specific calibration method. Secondly, a significant computational challenge arises from the iterative approximation process required to obtain the primary transmission $I_{\text{A}}^{p}$ in the Eq. \ref{FASKS}. \\

\subsubsection{FOSKS Module for Forward Scatter}

For clarity and conciseness, we detail the forward scatter correction process for scanning beam A, noting that the methodology for scanning beam B follows an identical computational framework.
We model the forward scatter $I_{\text{A}\leftarrow \text{A}}^{s}$ as the convolution of ambient scatter and cross-scatter corrected projection intensity $I_{\text{A}}^{\prime}$ with a forward scatter kernel $h_f$:
\begin{align}\label{OSKS}                                             I_{\textup{A}\leftarrow \textup{A}}^{s}\{k\}(u,v) = \iint\limits_{D_{\text{A}}} I_{\textup{A}}^{'}\{k\}(u^{\prime},v^{\prime})\, h_f\{k\}(u - u^{\prime},v - v^{\prime})\,\mathrm{d}u^{\prime} \mathrm{d}v^{\prime},
\end{align}
where forward scatter kernel $h_f$ is composed of a forward amplitude factor $c_f$ and a double Gaussian kernel $g_f$. The forward amplitude factor $c_f$ is defined as:
\begin{align}\label{c_f}
    c_f\{k\}(u^{\prime},v^{\prime}) = A\,\left(\frac{I^{'}_\textup{A}\{k\}(u^{\prime},v^{\prime})}{I^{*}(u^{\prime},v^{\prime})}\right)^{\alpha}\,\left(\ln\left(\frac{I^{*}(u^{\prime},v^{\prime})}{I^{'}_\textup{A}\{k\} (u^{\prime},v^{\prime})}\right)\right)^{\beta}.
\end{align}
The kernel function $g_f$ maintains the same formulation as described in Eq. \ref{guassKernel}.
Recognizing the computational complexity associated with direct convolution operations, we implement an efficient computational framework through the application of 2D FFT, which yields the following formula:
\begin{align}\label{FOSKS}
    I_{\textup{A}\leftarrow \textup{A}}^{s}\{k\}(u,v) =  \mathcal{F}^{-1}\left(\mathcal{F}\left( I_{\text{A}}^{'}\{k\}(u,v)\, c_f\{k\}(u,v)\right)\, \mathcal{F}\left(g_f(u,v)\right)\right).
\end{align}

The complete set of forward scatter parameters requiring calibration in Eq. \ref{FOSKS} is represented by the vector $\vec{\theta}_f = (A, B, \alpha, \beta,\sigma_1, \sigma_2) $.
To address the inherent limitations of the FASKS algorithm in parameter calibration, particularly in terms of accuracy and computational efficiency, we introduce a novel approach that leverages forward scatter prior information from MC simulations at sparse angles.

Specifically, GPU-based open-source MC simulation software MC-GPU \cite{badal2009accelerating} is employed for the simulation process due to its high computational efficiency. The reconstructed image  $\tilde{\mu}_{a,o}(\vec{x})$ is segmented into distinct material regions (e.g., air, water, and bone) based on their respective density values. These segmented templates are then used as input for the MC-GPU program, enabling accurate modeling of X-ray photon interactions within the scanned object.
The program outputs the primary transmission $I_{\text{MC}}^{p}$ and forward scatter $I_{\text{MC}}^{s}$, which are utilized to calibrate forward scatter parameters $\vec{\theta}_f$ for the scanned object within the FOSKS module. The total projection intensity $I_{\text{MC}}^{s}$ is defined as the sum of primary transmission and the forward scatter, i.e., $I_{\text{MC}}^{t} = I_{\text{MC}}^{p} + I_{\text{MC}}^{s}$. Subsequently, the forward scatter parameters $\vec{\theta}_f$ are calibrated by minimizing the least squares error between the forward scatter distribution estimated by the FOSKS module and the MC-simulated forward scatter label $I_{\text{MC}}^{s}$. The objective function $J_f$ is formulated as follows:
\begin{align}
    J_{f} &= \sum_{k = 1}^{K}\sum_{u = -H_u}^{H_u}\sum_{v = -H_v}^{H_v} \left(G\{k\}(u,v) - I_{\textup{MC}}^{s}\{k\}(u,v) \right)^2, \,\text{where} \\
     \quad G\{k\}(u,v) &= \mathcal{F}^{-1}\left(\mathcal{F}\left( I_{\text{MC}}^{t}\{k\}(u,v)\, c_f\{k\}(u,v)\right)\, \mathcal{F}\left(g_f(u,v)\right)\right) \nonumber,
\end{align}
where $K$ denotes the number of sparse sampling angles. 
Given the similarity of forward scatter distributions between adjacent scan angles, MC-GPU simulates X-ray photon interactions with the segmented templates only at sparse angles, significantly reducing computational time while maintaining high precision. Once the forward scatter parameters are calibrated, FOSKS module can be utilized to estimate forward scatter for any arbitrary scanning angle. 
Consequently, the estimated primary transmission of a dual-source imaging system $I_{\text{A}}^p$ is derived as:
\begin{align}
    I_{\text{A}}^{p}\{k\}(u,v) = I_{\text{A}}^{'}\{k\}(u,v) - \mathcal{F}^{-1}\left(\mathcal{F}\left( I_{\text{A}}^{'}\{k\}(u,v)\,c_f\{k\}(u,v)\right)\,\mathcal{F}\left(g_f\{k\}(u,v)\right)\right).
\end{align}
After correcting for ambient scatter obtained from measurements, cross-scatter derived from the xSKS module, and forward scatter computed by the FOSKS module, a range of analytical and iterative reconstruction algorithms\cite{feldkamp1984practical,wang2004ordered} can be utilized to reconstruct high-quality images free from scatter artifacts. Furthermore, to address potential beam-hardening artifacts, an iterative beam-hardening correction algorithm \cite{krumm2008reducing,trotta2022beam} may be incorporated, further enhancing the accuracy and reliability of the reconstructed images.

\section{Experiments}
This paper systematically evaluates the performance of the proposed method in correcting dual-source scatter artifacts through a comprehensive validation framework, combining multiple sets of MC simulation experiments and physical experiments. The MC simulation experiments are implemented using the open-source MC-GPU software, whereas the physical experiments are conducted on a micro-focus dual-source imaging system. To ensure consistency between simulation and practical imaging conditions, the geometric parameters in the simulation experiments are designed to closely match those of the dual-source system in our laboratory. To reduce computational complexity, the reconstruction following scatter correction in this study is limited to two dimensions. The scatter correction process is consistent across both scanning beams. For clarity, this paper presents the scatter correction results using scanning beam A as a representative example.
\subsection{Evaluation from Simulation}
\subsubsection{Experiments setups}
All relevant parameters for the Monte Carlo (MC) simulations are summarized in Table \ref{tab:MC_configurations}. 
Given the low-frequency characteristics inherent to scattering phenomena, the flat-panel detector array utilized in the MC simulations was downsampled from an initial resolution of $4608 \times 5890$ to a coarser resolution of $288 \times 368$ elements, significantly reducing the computational burden. As a result, the dimensions of each detector element were correspondingly scaled up from $0.00495 \times 0.00495 \,\text{cm}^2$ to $0.0792 \times 0.0792 \,\text{cm}^2$.

The X-ray energy spectrum employed in the MC simulations is generated from the Oxford Series6000 tube, operating at a tube voltage of 120 kVp. The spectrum is simulated using SpectrumGUI, an open-source X-ray spectrum simulation tool\cite{Spectrum}. \\

\begin{table}[htbp]
    \centering
    \caption{Scanning configurations of MC simulations}
    \label{tab:MC_configurations}
    \begin{tabular}{lcc}
        \toprule[1.5pt] 
        Configurations         & Source A &  Source B \\ 
        \midrule
        Voltage  (kVp)             & 120              & 120       \\
        Detector array        & $4608 \times 5890\,(288 \times 368)$    & -           \\
        Detector element size (cm) & $0.00495 \times 0.00495\,(0.0792 \times 0.0792 )$           & -            \\
        Source to axis distance (cm) & 26.71         & 29.84          \\
        Source to detector distance (cm) & 42.90        & 41.75         \\
        Number of projections           & 720               & 720        \\
        Histories per projection        & 3e9               & 3e9         \\
        \bottomrule[1.5pt] 
    \end{tabular}
\end{table}

A total of two sets of simulation experiments were conducted, with the complexity of the scatter distributions for the experimental phantoms increasing sequentially. 
The first experimental phantom is a multi-material water cylinder (0.83 g/$\textup{cm}^3$, 0.92 g/$\textup{cm}^3$, 1.00 g/$\textup{cm}^3$, 1.18 g/$\textup{cm}^3$) with a diameter of 12 cm and a height of 8 cm, incorporating a bone insert (1.92 g/$\textup{cm}^3$) of 1 cm diameter to form a dual-material water-bone phantom. 
The second experimental phantom is a cylindrical phantom designed in the shape of a Yin-Yang symbol, composed of water (1.00 g/$\textup{cm}^3$) and bone (1.92 g/$\textup{cm}^3$) materials, with a diameter of 9.6 cm and a height of 4 cm. Notably, this Yin-Yang simulation phantom is identical to the one utilized in the physical experiments, ensuring that the conclusions derived from the simulated and experimental results are mutually consistent and strictly validated. To minimize the interference of beam-hardening artifacts in the second set of experiments, a 1 mm copper filter was added to this MC simulation.
\subsubsection{Evaluation metric}
SPMAPE index is quantitatively evaluated in the projection domain to assess and compare the accuracy of various algorithms in estimating scatter distribution\cite{erath2021deep} .
\begin{align}
    \text{SPMAPE} &= \frac{1}{N} \sum \left| \frac{I_{\text{scatter,ref}} - I_{\text{scatter,est}}}{I_{\text{scatter,ref}}} \times \frac{I_{\text{scatter,ref}}}{I_{\text{prim}}} \right| \nonumber \\
                  &= \frac{1}{N} \sum \left| \frac{I_{\text{scatter,ref}} - I_{\text{scatter,est}}}{I_{\text{prim}}} \right|,
\end{align}
where $N$ represents the number of pixels, $I_{\text{scatter,ref}}$ denotes the reference of scatter, $I_{\text{scatter,est}}$ corresponds to the estimated scatter, $I_{\text{prim}}$ represents the ideal primary projection intensity. 
The root mean square error (RMSE) and mean absolute error (MAE) are used to evaluate and compare the quality of reconstructed images with different scatter correction algorithms.
\subsubsection{Results}
In each set of MC simulation experiments, we sequentially compared the performance of different methods in estimating cross-scatter, forward scatter, and total scatter in dual-source imaging systems. 

\begin{figure}[ht]
   \begin{center}
   \includegraphics[width=\textwidth]{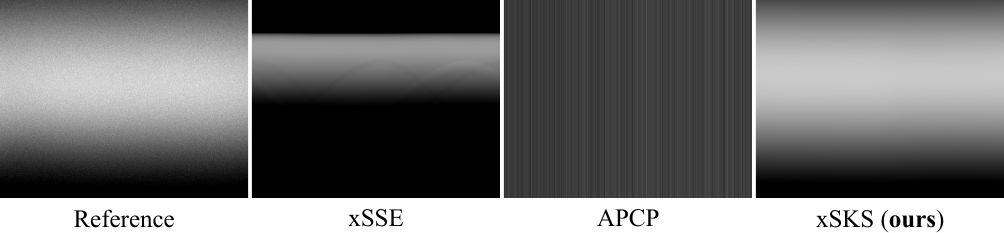}
   \caption{The cross-scatter sinograms estimated from the simulated water-bone phantom. The display window is [400, 1200].
   \label{C401_CrossProj}}  
    \end{center}
\end{figure}

\begin{figure}[ht]
   \begin{center}
   \includegraphics[width=\textwidth]{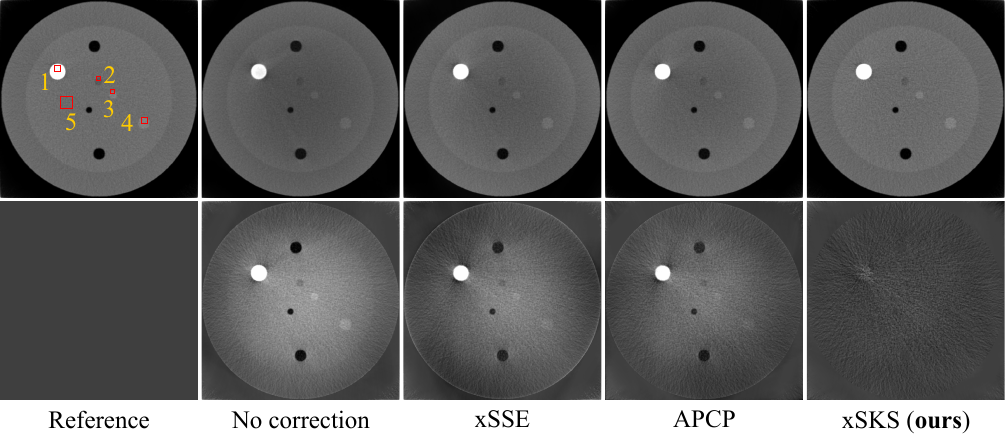}
   \caption
   {The cross-scatter corrected reconstructions of the simulated water-bone phantom. The display windows for reconstructions and error images are respectively [0, 0.6] and [-0.05, 0.15].
   \label{C401_CrossRecon} 
    }  
    \end{center}
\end{figure}

In the first experiment set using a water-bone phantom, Fig. \ref{C401_CrossProj} shows the cross-scatter estimates in the projection domain, while Fig. \ref{C401_CrossRecon} presents the corresponding reconstructed images with cross-scatter correction.
The reconstructed images and error maps in Fig. \ref{C401_CrossRecon} show that cross-scatter artifacts remain in the correction results of existing methods, whereas our proposed approach achieves reconstructed image that closely match the reference image, with no detectable cross-scatter artifacts. 

\begin{figure}[h]
   \begin{center}
   \includegraphics[width=\textwidth]{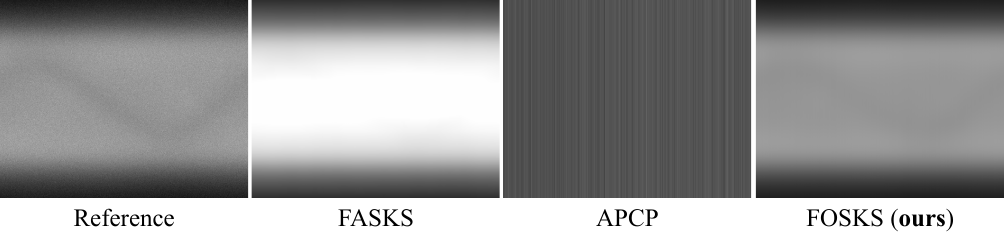}
   \caption
   {The forward scatter sinograms estimated from the simulated water-bone phantom. The display window is [300, 1200].
   \label{C401_ForwardProj} 
    }  
    \end{center}
\end{figure}

\begin{figure}[ht]
   \begin{center}
   \includegraphics[width=\textwidth]{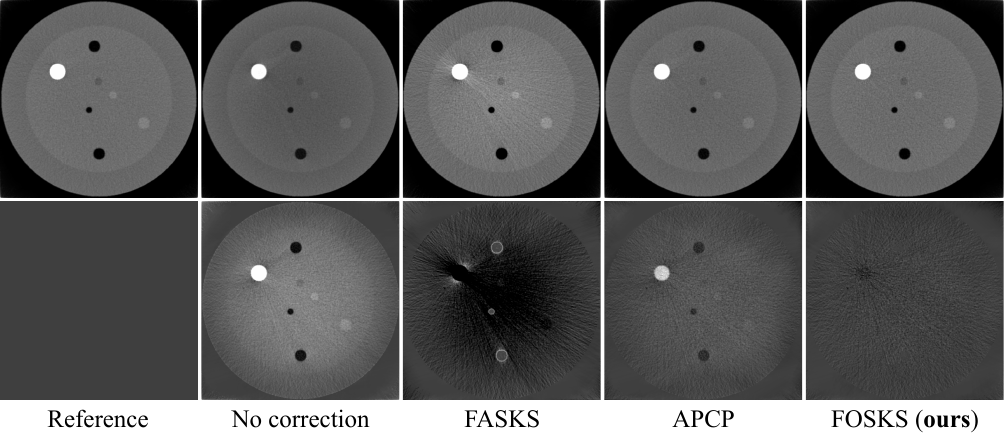}
   \caption
   {The forward scatter corrected reconstructions of the simulated water-bone phantom. The display windows for reconstructions and error images are respectively [0, 0.6] and [-0.05, 0.15].
   \label{C401_ForwardRecon} 
    }  
    \end{center}
\end{figure}

Fig. \ref{C401_ForwardProj} shows the forward scatter estimates in the projection domain, while Fig. \ref{C401_ForwardRecon} presents the corresponding reconstructed images with forward scatter correction.
The reconstructed images and error maps in Fig. \ref{C401_ForwardRecon} reveal that the FASKS algorithm tends to overestimate forward scatter in high-attenuation bone regions, and the APCP algorithm still exhibits residual forward scatter artifacts. In contrast, the proposed method generates reconstructed results where the high-attenuation bone regions are slightly higher than the reference values, while the low-density regions align closely with the reference values. 

\begin{figure}[h!]
   \begin{center}
   \includegraphics[width=\textwidth]{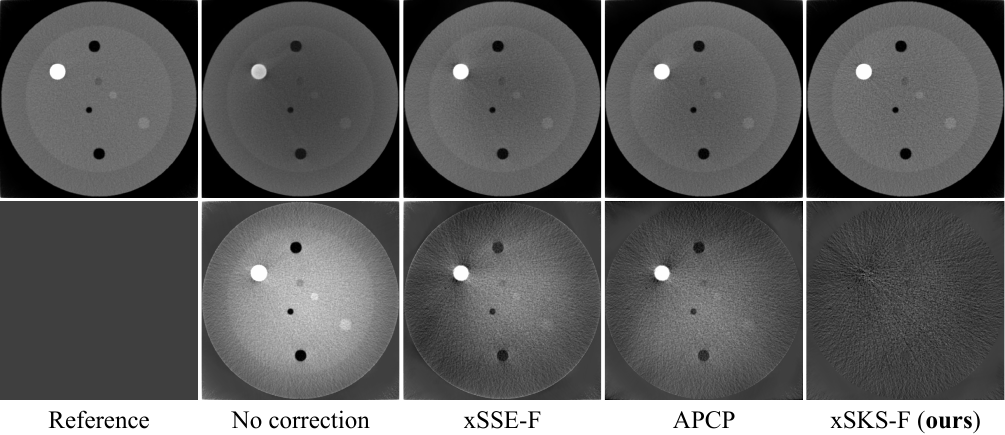}
   \caption
   {The cross-scatter and forward scatter corrected reconstructions of the simulated water-bone phantom. The display windows for reconstruction results and error images are respectively [0, 0.6] and [-0.05, 0.15].
   \label{C401_TotalRecon} 
    }  
    \end{center}
\end{figure}
The reconstructed images and corresponding error maps in Fig. \ref{C401_TotalRecon}, obtained after simultaneous correction of cross-scatter and forward scatter, further validate the superiority of the proposed scatter correction model.
Table \ref{Quantitative Results} quantitatively compares the scatter distributions estimated by different methods using the SPMAPE metric, along with the RMSE and MAE metrics of the reconstructed images with the corresponding scatter correction. 
Moreover, Table \ref{ROI Index} summarizes the averaged RMSE metrics of the scatter corrected reconstructions within the five regions of interest (ROI), outlined by red rectangles in Fig. \ref{C401_CrossRecon}.

\begin{figure}[h]
   \begin{center}
  \includegraphics[width=\textwidth]{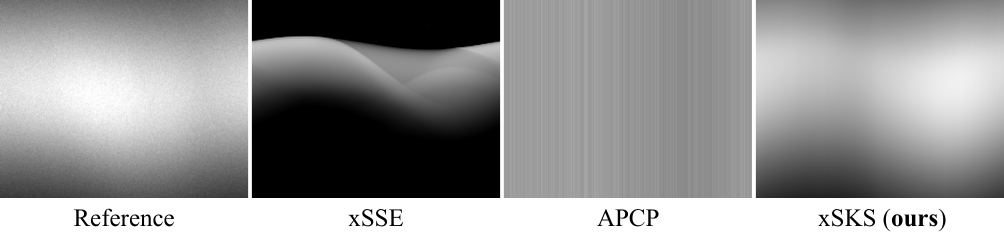}
   \caption
   {The cross-scatter sinogram results estimated from the simulated Yin-Yang phantom. The display window is [300,2000]. 
   \label{Yin-Yang_CrossProj} 
    }  
    \end{center}
\end{figure}

\begin{figure}[ht]
   \begin{center}
  \includegraphics[width=\textwidth]{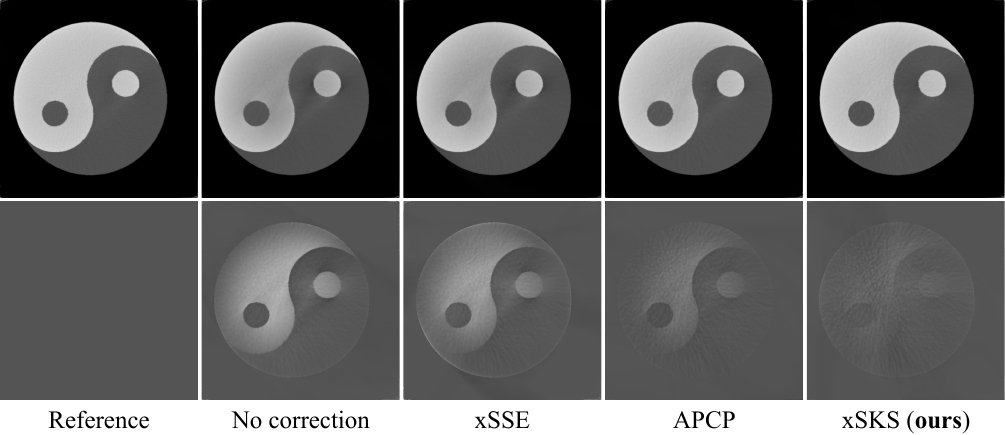}
   \caption
   {The cross-scatter correction reconstruction results from the simulated Yin-Yang phantom. The display windows for reconstruction results and error images are respectively [0, 0.6] and [-0.1, 0.2].
   \label{Yin-Yang_CrossRecon} 
    }  
    \end{center}
\end{figure}

In the second experiment set using a Yin-Yang phantom, Fig. \ref{Yin-Yang_CrossProj} shows the cross-scatter estimates in the projection domain.
Consistent with the observations in Fig. \ref{C401_CrossProj}, the xSSE algorithm solely accounts for cross-scatter along the primary paths, leading to a underestimation of cross-scatter intensity.
During the MC simulation experiments conducted without detector cross-scatter, the APCP algorithm approximates cross-scatter as a constant for each projection angle. Despite the fact that the magnification ratio is not significant, the distribution of cross-scatter exhibits smooth properties. As a result, this algorithm is able to achieve a certain level of cross-scatter correction efficacy, even though its modeling process is relatively simplified. 
The error maps of the reconstructed results in Fig. \ref{Yin-Yang_CrossRecon} indicate that the proposed
cross-scatter correction module delivers the optimal correction performance, exhibiting only minimal residual artifacts at the water-bone interfaces. \\


\begin{figure}[h]
   \begin{center}
  \includegraphics[width=\textwidth]{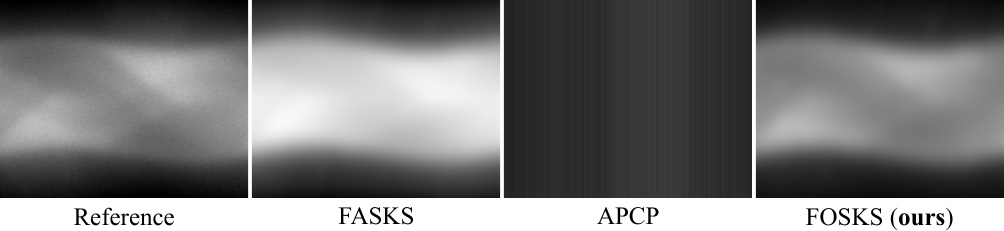}
   \caption
   {The forward scatter sinogram results estimated from the simulated Yin-Yang phantom. The display window is [800,3200].
   \label{Yin-Yang_ForwardProj} 
    }  
    \end{center}
\end{figure}
\begin{figure}[h!]
   \begin{center}
   \includegraphics[width=\textwidth]{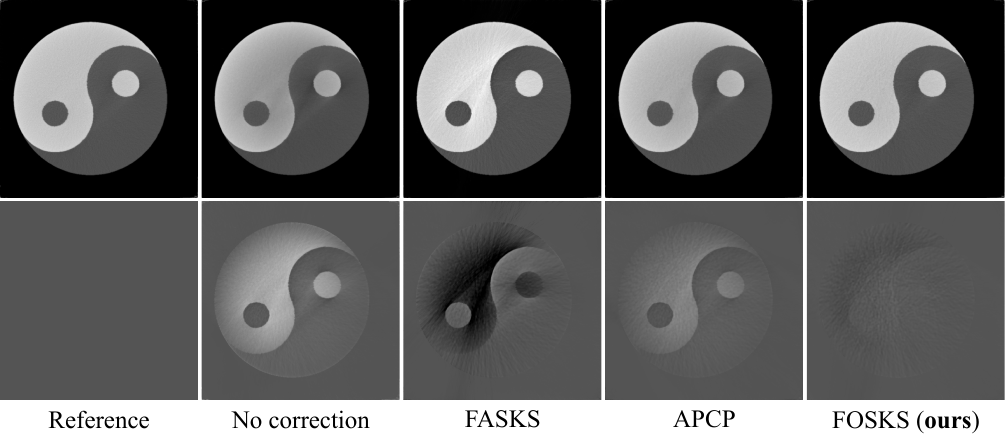}
   \caption
   {The forward scatter correction reconstruction results from the simulated Yin-Yang phantom. The display windows for reconstruction results and error images are respectively [0, 0.6] and [-0.1, 0.2].
   \label{Yin-Yang_ForwardRecon} 
    }  
    \end{center}
\end{figure}
\begin{figure}[h]
   \begin{center}
  \includegraphics[width=\textwidth]{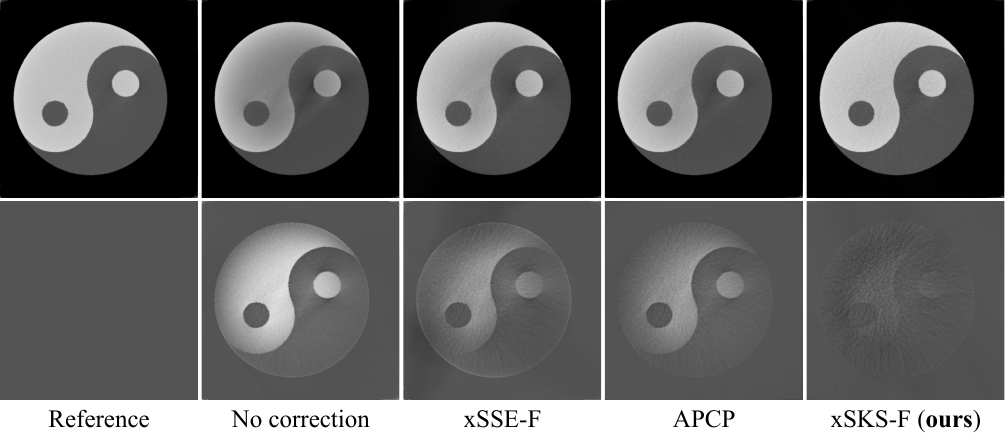}
   \caption
   {The cross-scatter and forward scatter correction reconstruction results from the simulated Yin-Yang phantom. The display windows for reconstruction results and error images are respectively [0, 0.6] and [-0.1, 0.2].
   \label{Yin-Yang_TotalRecon} 
    }  
    \end{center}
\end{figure}

In contrast to cross-scatter, forward scatter exhibits a stronger dependence on the shape and material composition of the object. As illustrated in Fig. \ref{Yin-Yang_ForwardProj} and Fig. \ref{Yin-Yang_ForwardRecon}, the APCP algorithm, which approximates forward scatter as a projection angle-dependent constant, may lead to reduced accuracy in forward scatter correction. On the other hand, the FASKS algorithm successfully estimates the overall shape of the forward scatter distribution. Nevertheless, as highlighted in the methodology section, this approach tends to overestimate 
forward scatter in regions adjacent to high-attenuation materials. 
Our proposed method leverages forward scatter priors derived from MC simulations at sparsely sampled angles to calibrate the parameters of the FOSKS module. The reconstructed error map demonstrates that our method effectively eliminates forward scatter artifacts, achieving superior correction performance. The reconstructed images and corresponding error maps in Fig. \ref{Yin-Yang_TotalRecon}, obtained after simultaneous correction of cross-scatter and forward scatter. \\

 \begin{table}[h!] 
    \centering
    \caption{Averaged Linear Attenuation Coefficient ($\textup{cm}^{-1}$) and RMSE in ROI}
    \label{ROI Index}
    \begin{tabular}{llcccccc}
        \toprule[1.5pt] 
        Scatter & Methods       &1      & 2     & 3     & 4     & 5     &$\text{RMSE}_{\text{ROI}}$ \\
        \midrule
                & Reference     & 0.9148& 0.2145& 0.2898& 0.3032& 0.2545& 0 \\
        \midrule [0.1pt]                
        Cross   & No correction & 0.5654&0.1668 & 0.2094& 0.2384& 0.1896& 0.1669 \\
                & xSSE   & 0.6997   & 0.1836  & 0.2385   & 0.2704  & 0.2047  &0.1034 \\
                & APCP   & 0.8124 & 0.1961 & 0.2618 & 0.2897 & 0.2258  &0.0502\\                
                & xSKS(\textbf{ours})  & \textbf{0.9037}   & \textbf{0.2161}   & \textbf{0.2901} & \textbf{0.3002}  & \textbf{0.2516} & \textbf{0.0054} \\
        \midrule   
        Forward & No correction & 0.6085 & 0.1780 & 0.2245 & 0.2438 & 0.2073 & 0.1450\\
                & FASKS & 0.8055 & 0.2508 & 0.3640 & 0.3446 & 0.2982 & 0.0669\\
                & APCP & 0.8055 & 0.2039 & 0.2661 & 0.2801 & 0.2407 & 0.0517\\
                & FOSKS(\textbf{ours}) & \textbf{0.9284} & \textbf{0.2155} &\textbf{0.2905} & \textbf{0.3023} & \textbf{0.2558} & \textbf{0.0061}\\
                
        \midrule    
        Total   & No correction   & 0.4464   & 0.1465  & 0.1761  & 0.2000 &0.1664& 0.2260 \\ 
                & xSSE-F & 0.6945 & 0.1717 & 0.2310 & 0.2747 & 0.2093 & 0.1065\\
          	& APCP  & 0.7226   & 0.1846  & 0.2390  & 0.2662  &0.2177 & 0.0929  \\
                & xSKS-F(\textbf{ours}) & \textbf{0.9119}   & \textbf{0.2139}   & \textbf{0.2855} & \textbf{0.2967}  & \textbf{0.2486} & \textbf{0.0046} \\
        \bottomrule[1.5pt] 
    \end{tabular}
   
    \vspace*{2pt} 
    \raggedright\footnotesize\textsuperscript{*}ROI 1--5 correspond to distinct regions of interest in the simulated water-bone phantom.
\end{table} 
\begin{table}[h] 
    \centering
    \caption{Quantitative Results of Simulated Phantoms}
    \label{Quantitative Results}
	\setlength{\tabcolsep}{4pt} 
    \begin{tabular}{llcccccc}
        \toprule
         Scatter    & Methods      & \multicolumn{3}{c}{Water-Bone phantom} & \multicolumn{3}{c}{Yin-Yang phantom} \\ 
                 \cmidrule(lr){3-5} \cmidrule(lr){6-8} 
                 &        & SPMAPE   & RMSE  & MAE    & SPMAPE   & RMSE  & MAE  \\
        \midrule
  Cross  & No correction & 0.3095 & 0.0424 & 0.0300 & 0.1732 & 0.0311 & 0.0157 \\
         & xSSE & 0.1534 & 0.0269 & 0.0178 & 0.1021 & 0.0226 & 0.0123 \\
         & APCP & 0.0615 & 0.0147 & 0.0100 & 0.0423 & 0.0109 & 0.0054 \\
         & xSKS(\textbf{ours}) & \textbf{0.0108} & \textbf{0.0067} & \textbf{0.0050} & \textbf{0.0197} & \textbf{0.0045} & \textbf{0.0028} \\
        \midrule
    Forward & No correction & 0.2710 & 0.0369 & 0.0268 & 0.2042 & 0.0329 & 0.0174 \\ 
    & FASKS & 0.1119 & 0.0326 & 0.0175 & 0.0749 & 0.0241 & 0.0117 \\
    & APCP & 0.0762 & 0.0136 & 0.0093 & 0.0786 & 0.0158 & 0.0083 \\
     & FOSKS(\textbf{ours}) & \textbf{0.0079} & \textbf{0.0063} & \textbf{0.0046} & \textbf{0.0089} & \textbf{0.0038} & \textbf{0.0025} \\
     \midrule
    Total  & No correction & 0.5805 & 0.0638 & 0.0484 & 0.3774 & 0.0530 & 0.0290 \\ 
    & xSSE-F & 0.1531 & 0.0284 & 0.0187 & 0.1011 & 0.0199 & 0.0107 \\
    & APCP   & 0.1299 & 0.0245 & 0.0163 & 0.1206 & 0.0246 & 0.0129 \\
    & xSKS-F (\textbf{ours}) & \textbf{0.0132} & \textbf{0.0091} & \textbf{0.0067} & \textbf{0.0200} & \textbf{0.0055} & \textbf{0.0039} \\
    \bottomrule
    \end{tabular}
\end{table}

\clearpage
\newpage
\subsection{Physical Experiments}
\subsubsection{Experiments setups}
\begin{figure}[H]
   \begin{center}
   \includegraphics[width=\textwidth]{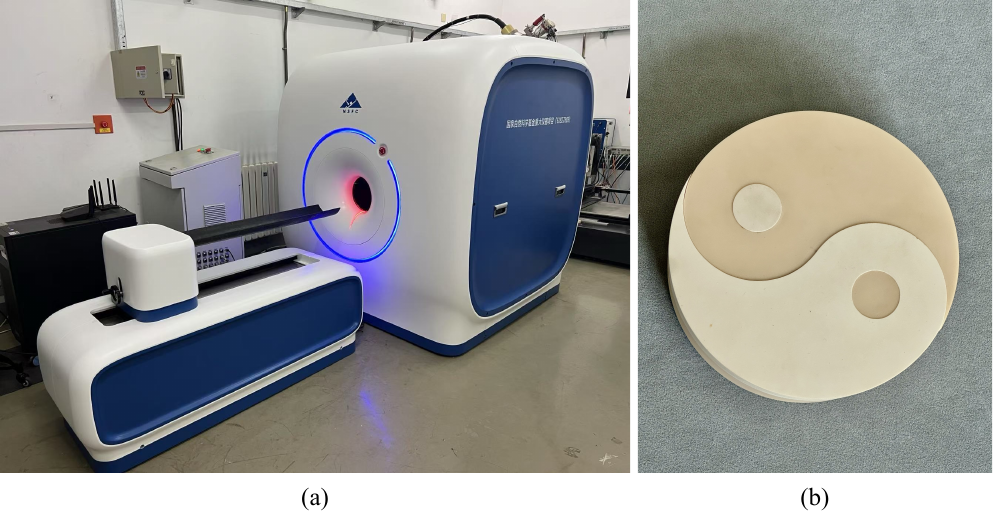}
   \caption
   { (a) The micro-focus dual-source imaging system. (b) The Yin-Yang phantom.
   \label{Equipment} 
    }  
    \end{center}
\end{figure}

As illustrated in the Fig \ref{Equipment} , the physical experiment was conducted on our self-developed dual-source imaging system, which features two approximately perpendicular scanning beams. Scanning beam A consists of a standard X-ray source A (L9181, Hamamatsu, Japan) coupled with an energy-integrating detector A (Xineos-2329, DALSA, Canada). Detector A comprises $4608 \times 5890$ detector elements, each with a pixel size of $0.00495 \times 0.00495 \, \text{cm}^2$. By binning $2 \times 2$ detector elements, the raw projection data is obtained with a resolution of $2304 \times 2945$ pixels. Scanning beam B is equipped with a standard X-ray source B (Libra13UINE, iRay, China) and a photon-counting detector B (EIGER2, DECTRIS, Switzerland). Detector B consists of $1024 \times 2068$ detector elements, each measuring $0.0075 \times 0.0075 \, \text{cm}^2$. 

To validate the effectiveness of our proposed scatter model, a custom-designed Yin-Yang phantom with a diameter of $9.6  \,\text{cm} $ and a height of $4 \text{cm} $ was utilized. The specific scanning parameters are summarized in Table \ref{tab:scanning_configurations}, while the detailed scanning procedure is outlined in Table \ref{Scanning modes}.\\

\begin{table}[ht] 
    \centering
    \caption{Scanning configurations of physical experiments}
    \label{tab:scanning_configurations}
    \begin{tabular}{lcc}
        \toprule[1.5pt] 
        Configurations         & Source A &  Source B \\ 
        \midrule
        Voltage (kVp)/ Filter                & 120 / No       & 120  / No       \\
        Current (mA)                       & 300             & 300             \\
        Exposure time per projection (s)   & 0.2             & 0.2             \\
        Source to object distance (cm)     & 26.71          & 29.84          \\
        Source to detector distance (cm)   & 42.90         & 41.75        \\
        Number of projections              & 720            & 720               \\
        \bottomrule[1.5pt] 
    \end{tabular}
\end{table} 
\ \\
\begin{table}[h]
    \centering
    \caption{Scanning protocols for scatter correction}
    \label{Scanning modes}
    \begin{tabular}{l}
        \toprule[1.5pt] 
        1. Data acquisition without object in single-source mode.       \\ 
	   \quad  (a)  incident intensity $I_{\text{A}}^{*}$, the detector A acquisition with beam A, \\ 
         \quad  (b)  ambient scatter $I_{\text{A}\leftarrow {\text{B}}}^{*}$, the detector A acquisition from source B. \\ 
          \midrule [0.1pt]
        2. Data acquisition with object in single-source mode.  \\ 
          \quad  (a)  output intensity $I_{\text{A}}$, the sinogram of beam A, \\ 
           \quad (b)  sum of ambient and cross scatter $I_{\text{A}\leftarrow {\text{B}}}^{t,s}$, the detector A acquisition from source B. \\ 
           \midrule [0.1pt]
        3. Data acquisition with object in dual-source mode. \\
          \quad  (a)   output intensity $I_{\text{A}}^{t}$, the sinogram of beam A.  \\ 
       \bottomrule[1.5pt] 
    \end{tabular} 
\end{table}
\subsubsection{Results}
\begin{figure}[h]
   \begin{center}
    \includegraphics[width=\textwidth]{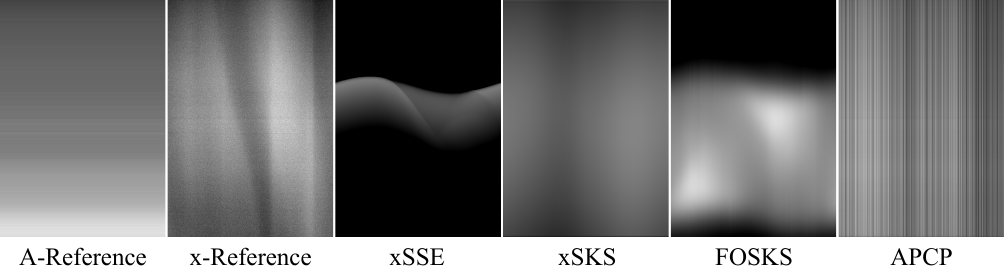}
   \caption
   {Different scatter sinogram results from the real Yin-Yang phantom. A-Reference:  the reference of measured ambient scatter; x-Reference: the reference of measured cross-scatter. APCP estimates the total scatter and displays the result in the range of [0, 1800], while the display window for other images is [200, 600]. 
   \label{Real_YinYang_CrossProj} 
    }  
    \end{center}
\end{figure}

Based on the acquisition protocol outlined in Table \ref{Scanning modes}, the corresponding ambient scatter $I_{\text{A}\leftarrow {\text{B}}}^{*}$ was measured, as depicted in A-Reference in Fig. \ref{Real_YinYang_CrossProj}. Subsequently, the $I_{\text{A}\leftarrow {\text{B}}}^{*}$ was subtracted from $I_{\text{A}\leftarrow {\text{B}}}^{t,s}$ to derive the cross-scatter label x-Reference, which serves as the ground truth for the modeling approach. As demonstrated in Fig. \ref{Real_YinYang_CrossProj}, the cross-scatter estimated by the proposed xSKS module shows significantly better agreement with the x-Reference label compared to the existing xSSE algorithm. 

\begin{figure}[h]
   \begin{center}
    \includegraphics[width=\textwidth]{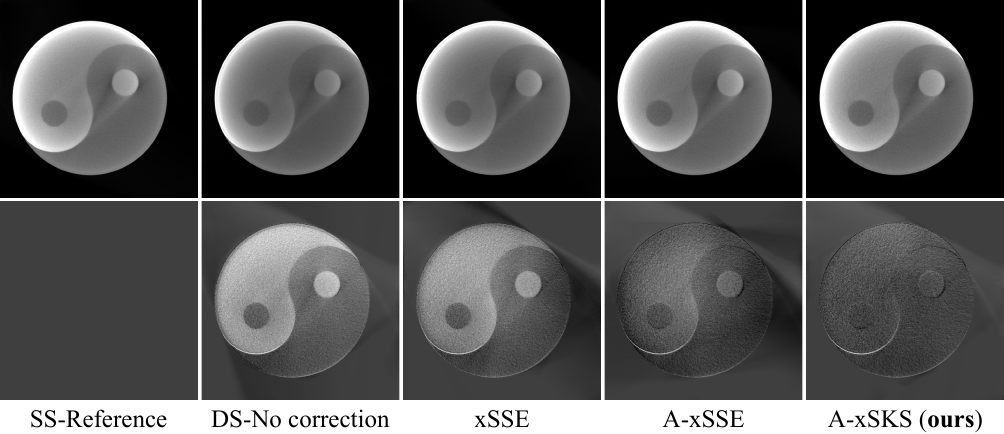}
   \caption
   {Reconstructed images with ambient scatter and cross-scatter corrections. SS-Reference: the single-source reference; DS-No correction: dual-source scan without scatter correction; A-xSSE and A-xSKS: dual-source scan with ambient scatter and xSSE and xSKS cross-scatter corrections, respectively.
   The display window for the reconstructed images is set to [0, 0.65], while the error maps are displayed in the range of [-0.05, 0.15].
   \label{Real_YinYang_CrossRecon} 
    }  
    \end{center}
\end{figure}

As illustrated in Fig. \ref{Real_YinYang_CrossRecon}, in comparison to the uncorrected reconstruction, the xSSE algorithm, which models cross-scatter along the primary paths, achieves a moderate reduction in scatter artifacts within the reconstructed images. Furthermore, the A-xSSE algorithm builds upon the xSSE algorithm by integrating ambient scatter correction, enhancing the suppression of scatter artifacts, particularly in high-attenuation bone regions. Notably, our proposed A-xSKS algorithm not only incorporates ambient scatter correction but also extends the existing xSSE algorithm to include cross-scatter contributions beyond the primary paths, significantly reducing scatter artifacts in the reconstructed images. The quantitative results presented in Table \ref{tab:SPMAPE_MAE comparison} further validate the superior accuracy and effectiveness of the proposed method in scatter modeling and estimation. \\

\begin{figure}[ht]
   \begin{center}
    \includegraphics[width=\textwidth]{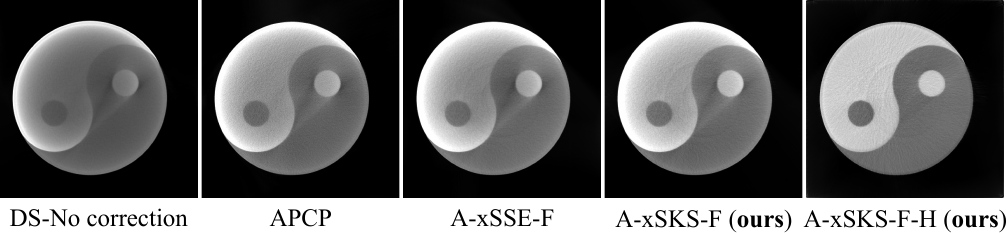}
   \caption
   {Reconstructed images with scatter and further harden corrections. A-xSSE-F and A-xSKS-F: dual-source scan with ambient scatter, xSSE and xSKS cross-scatter respectively, and FOSKS forward scatter corrections; A-xSKS-F-H: A-xSKS-F with additional beam hardening correction. The display window for the reconstructed images is set to [0, 0.65], while the error maps are displayed in the range of [-0.05, 0.15].
   \label{Real_YinYang_TotalRecon} 
    }  
    \end{center}
\end{figure}

The reconstructed images in Fig. \ref{Real_YinYang_CrossRecon} demonstrate the residual forward scatter artifacts, which degrade the reconstruction quality. To address this issue, the proposed FOSKS module was employed for forward scatter estimation, as illustrated in Fig. \ref{Real_YinYang_CrossProj}. 
The effectiveness of this module in accurately estimating forward scatter has been rigorously validated through corresponding simulation experiments. 
Following the application of the FOSKS module for further forward scatter correction, the reconstructed images obtained using the A-xSSE-F and our A-xSKS-F model exhibit significant improvement, as illustrated in Fig. \ref{Real_YinYang_TotalRecon}.
For comparative analysis, the APCP method was also adopted as a baseline to evaluate the performance of our model. The experimental results in Fig. \ref{Real_YinYang_TotalRecon} demonstrate that the proposed A-xSKS-F model achieves superior performance in scatter correction compared to other methods, particularly in regions with severe scatter artifacts, such as high-attenuation bone areas and the central regions of the object.
Since no additional filter was employed during the acquisition process, the reconstructed result with A-xSKS-F model, while demonstrating excellent scatter correction, still exhibit residual beam-hardening artifacts. To mitigate this issue, a beam-hardening correction was integrated into the proposed method, yielding the enhanced A-xSKS-F-H approach. \\

\begin{table}[ht]
    \centering
    \caption{Quantitative analysis for reconstructions in Fig. \ref{Real_YinYang_CrossRecon}}
    \label{tab:SPMAPE_MAE comparison}
    \begin{tabular}{lcccccc}
        \toprule
         {Metric} & {DS-No correction} & {xSSE} & {A-xSSE} & {A-xSKS (\textbf{ours})} \\
        \midrule
        RMSE & 0.0441 & 0.0349 & 0.0188  & \textbf{0.0136}  \\
        MAE  & 0.0278 & 0.0173 &  0.0134 & \textbf{0.0083}  \\
        \bottomrule
    \end{tabular}
\end{table}


\section{Conclusion}
In this study, we proposed a comprehensive scatter correction model for micro-foucs dual-source imaging systems, addressing three key scatter components: ambient scatter, cross-scatter, and forward scatter. By modeling ambient scatter as a discrete point source and measuring it during air-scans, we demonstrated its negligible angular variation impact on reconstruction accuracy.
The proposed xSKS module significantly enhanced cross-scatter estimation by incorporating single and multiple cross-scatter events along non-primary paths, which were previously neglected in the xSSE algorithm. 
Furthermore, the FOSKS module extended the FASKS algorithm to correct forward scatter, offering a more efficient and practical parameter calibration approach.

In the current model, it is assumed that the distribution of ambient scatter is dependent on the hardware configuration during the scanning process and remains independent of the scanned object. Whether ambient scatter undergoes variations during scans involving the object necessitates further investigation through extensive experimental measurements.
Furthermore, the cross-scatter parameters in the xSKS module are derived from cross-scatter labels selected at a single angle, which determines the estimated cross-scatter sinogram.
Preliminary experimental results indicate that the estimated cross-scatter sinogram exhibits minimal sensitivity to the selection of the cross-scatter label angle.


%
 

\clearpage

\section*{References}
\addcontentsline{toc}{section}{\numberline{}References}
\vspace*{-15mm}






\bibliographystyle{./medphy.bst}    


\end{document}